\documentclass[aps,preprint,nofootinbib,preprintnumbers,eqsecnum]{revtex4}

\usepackage{color}

\usepackage[
      colorlinks=true,
      linkcolor=blue,
      urlcolor=blue,
      filecolor=black,
      citecolor=red,
      pdfstartview=FitV,
      pdftitle={},
        pdfauthor={Thomas Faulkner, Hong Liu, Mukund Rangamani},
        pdfsubject={},
        pdfkeywords={},
        pdfpagemode=None,
        bookmarksopen=true
      ]{hyperref}

\usepackage[normalem]{ulem}
\usepackage{amsmath}
\usepackage{enumerate}
\usepackage{amsfonts}
\usepackage{yfonts}

\usepackage{subfigure}
\usepackage{psfrag}

\usepackage{epsfig}
\usepackage[latin1]{inputenc}
\usepackage{float}
\usepackage{graphicx}
\usepackage{cancel}
\usepackage{mathrsfs}
\usepackage{amssymb}
\usepackage{amsfonts}
\usepackage{amsmath}
\usepackage{slashed}

\usepackage{graphicx}
\usepackage{bm}

\def\ie{{\it i.e.}}

\def\({\left(}
\def\){\right)}
\def\[{\left[}
\def\]{\right]}
\def\<{\langle}
\def\>{\rangle}

\newcommand\half{{\ensuremath{\frac{1}{2}}}}

\newcommand\field[1]{{\ensuremath{\mathbb{{#1}}}}}

\newcommand\vev[1]{{\ensuremath{\left\langle{#1}\right\rangle}}}

\newcommand{\RR}{\field{R}}

\newcommand{\be}{\begin{equation}}
\newcommand{\ee}{\end{equation}}
\newcommand{\bea}{\begin{eqnarray}}
\newcommand{\eea}{\end{eqnarray}}
\newcommand{\bwt}{\begin{widetext}}
\newcommand{\ewt}{\end{widetext}}

\newcommand{\bi}{\begin{itemize}}
\newcommand{\ei}{\end{itemize}}
\newcommand{\ben}{\begin{enumerate}}
\newcommand{\een}{\end{enumerate}}
\newcommand{\bca}{\begin{cases}}
\newcommand{\eca}{\end{cases}}
\newcommand{\bln}{\begin{align}}
\newcommand{\eln}{\end{align}}
\newcommand{\bst}{\begin{split}}
\newcommand{\est}{\end{split}}

\newcommand\ep{\epsilon}

\newcommand\Sig{\Sigma}
\newcommand\lam{\lambda}

\newcommand\om{\omega}

\newcommand\Ga{{\ensuremath{{\Gamma}}}}
\newcommand\de{{\ensuremath{{\delta}}}}
\newcommand\De{{\ensuremath{{\Delta}}}}

\def\th{{\theta}}

\newcommand\ov{\over}
\newcommand\ha{{\half}}

\def\le{\left}
\def\ri{\right}

\newcommand\sG{{\ensuremath{{\mathcal G}}}}

\newcommand\sN{{\ensuremath{{\mathcal N}}}}
\newcommand\sO{{\ensuremath{{\mathcal O}}}}
\newcommand\sR{{\ensuremath{{\mathcal R}}}}

\renewcommand{\Im}{\textrm{Im}\,}

\newcommand{\vk}{{\vec k}}

\newcommand{\ircft}{{eCFT$_1$}}
\newcommand{\slql}{{semi-local quantum liquid}}
\newcommand{\Slql}{{SLQL}}

\newcommand{\nb}[1]{#1}

\begin{document}

\title {
Semi-local quantum liquids}

\preprint{MIT-CTP 4267}

\author{Nabil Iqbal}
\affiliation{Center for Theoretical Physics, Massachusetts Institute of Technology,
Cambridge, MA 02139 }
\author{ Hong Liu}
\affiliation{Center for Theoretical Physics,
Massachusetts
Institute of Technology,
Cambridge, MA 02139 }
\author{M\'ark Mezei}
\affiliation{Center for Theoretical Physics, Massachusetts Institute of Technology,
Cambridge, MA 02139 }

\begin{abstract}

Gauge/gravity duality applied to strongly interacting systems at finite density predicts a universal {\it intermediate energy} 
phase to which we refer as a {\it semi-local quantum liquid}. Such a phase is characterized by a finite spatial correlation length, but an infinite correlation {\it time} and associated nontrivial scaling behavior in the time direction, as well as a nonzero entropy density. For a holographic system at a nonzero chemical potential, this unstable phase 
sets in at an energy scale of order of the chemical potential, and orders at lower energies into other 
phases;  examples include superconductors, and antiferromagnetic-type states. In this paper we give examples in which it also orders into Fermi liquids of ``heavy'' fermions.
While the precise nature of the lower energy state depends on the specific dynamics of the individual system, we argue that the semi-local quantum liquid emerges universally at intermediate energies through {\it deconfinement} (or equivalently {\it fractionalization}). We also discuss the possible relevance of such a semi-local quantum liquid to heavy electron systems and the strange metal phase of  high temperature cuprate superconductors.

\end{abstract}

\today

\maketitle



\section{Introduction}

Understanding phases of matter for which there is no quasiparticle description 
presents some of the most challenging problems in physics. 
 In condensed matter physics prominent examples include the
  ``strange metals'' occurring in the normal state of the high temperature cuprate superconductors, 
and heavy electron systems near a quantum phase transition. In the strange metallic phase, 
electronic excitations near a Fermi surface have a large decay rate and as a result individual
electrons lose their integrity. There has been an accumulation of examples, but so far no satisfactory theoretical framework exists to describe them. Finding solvable examples of quantum phases with no quasiparticles should provide a valuable guide for the search for such a framework. 

During the last decade, developments in string theory have revealed surprising and profound
connections between gravity and many-body systems.  The so-called gauge/gravity duality, relates  a classical gravity theory in a weakly curved $(d + 1)$-dimensional anti-de Sitter (AdS$_{d+1}$) spacetime to a strongly-coupled $d$-dimensional quantum
field theory living on its boundary~\cite{Maldacena:1997re,Witten:1998zw,Gubser:1998bc}. In particular, black holes have played a universal role in characterizing quantum phases without quasiparticles, giving important insight into dynamical behavior of such systems. 

For a strongly interacting system (in $d$ spacetime dimensions) at finite density and zero temperature, an extremal charged black hole, whose near horizon region is given by AdS$_2 \times \RR^{d-1}$, 
provides the simplest gravity description~\cite{Chamblin:1999tk} and has played an important role
in recent explorations.  
In this paper we \nb{attempt to} further clarify the boundary theory interpretation of the gravity in AdS$_2 \times \RR^{d-1}$.  
We argue that it represents a universal fractionalized intermediate-energy phase, to which we refer as a {\it semi-local quantum liquid}. Such a phase does not have a quasiparticle description, and is characterized by a finite spatial correlation length, but an infinite correlation {\it time} and associated nontrivial scaling behavior in the time direction, as well as a nonzero entropy density. 
For a holographic system at a nonzero chemical potential, this unstable phase 
sets in at an energy scale of order of the chemical potential, and then orders into other 
phases at lower energies, with previously known examples including holographic superconductors~\cite{Hartnoll:2009sz,herzogr,Horowitz:2010gk}, antiferromagnetic-type states~\cite{Iqbal:2010eh}
and certain spatially modulated phase~\cite{Nakamura:2009tf}. We provide examples in which a lower energy phase is given by a Fermi liquid of ``heavy fermions,'' based on an earlier observation~\cite{Hartnoll:2009ns} on fermionic instabilities of AdS$_2 \times \RR^{d-1}$. While the precise nature of the lower energy state depends on the specific dynamics of the individual system, we argue that the semi-local quantum liquid arises universally from these lower energy phases through deconfinement~(or in condensed matter language, fractionalization). 

The plan of the paper is as follows. In Sec.~\ref{sec:uni} we introduce the notation of an intermediate-energy semi-local quantum liquid phase. In Sec.~\ref{sec:fermi} we give examples in which a semi-local quantum liquid settles into a Fermi liquid phase of heavy fermions 
at low energies. In Sec.~\ref{sec:sc} we discuss the underlying physical mechanism through which a semi-local quantum liquid  makes transition to a  superconducting state. In Sec.~\ref{sec:con} we conclude with a discussion of the possible relevance of the findings here for the strange metal phase of cuprate superconductors and heavy electron systems. 


\section{Universal semi-local quantum liquids} \label{sec:uni}

 Many examples of field theories with gravity duals are now known in various spacetime dimensions.  Well-studied examples include $\sN=4$ Super-Yang-Mills theory in $d=4$, and ABJM theory in $d=3$~\cite{Bagger:2007vi,Gustavsson:2007vu,Aharony:2008ug}. These theories essentially consist of elementary bosons and fermions interacting with
non-Abelian gauge fields. At a heuristic level one may visualize such a theory as 
the continuum limit of a lattice system where the number of degrees of freedom at each lattice site is of order $O(N^2)$, with $N$ the rank of the gauge group. 
The classical gravity approximation in the bulk corresponds to the strong coupling regime and the large $N$ limit, where the Newton's 
constant $G_N \propto {1 \ov N^2}$.  In addition to these theories, there also exist vastly many asymptotically-AdS vacua of string theory, each of which is believed to give rise to an example of the correspondence,
though an explicit description of the dual field theory is not known for most vacua.

In our discussion below we will take the so-called ``bottom-up'' approach, postulating a certain type of operator spectrum without referring to a specific theory. This approach is suitable for our current purposes as we are interested in extracting the possibly  universal features of a class of systems rather than understanding the detailed phase structure of any particular system. While our discussion applies to both $d=3$ and $d=4$, for definiteness we restrict to  $d=3$ and  take the field theory to be one with a conformally invariant vacuum, which amounts to working with gravity in an asymptotic AdS$_4$ spacetime. This conformal symmetry will not play a role in our results below as it will be broken by putting the system at a finite chemical potential.

For a boundary system with a conserved $U(1)$ charge a {\it universal} sector on the gravity side which is common to many theories with a gravity dual is the Einstein-Maxwell 
system with the gravitational field mapped to the boundary theory stress tensor and 
the Maxwell field $A_M$ mapped to the conserved $U(1)$ current  $J_\mu$.
At a  finite chemical potential $\mu$ for the $U(1)$, the Einstein-Maxwell system is described by a charged black hole whose horizon is topologically $\RR^2$ 
in AdS$_4$ spacetime~\cite{Chamblin:1999tk}.  In the zero temperature limit, the charged black hole develops a degenerate horizon and near the horizon the spacetime geometry is 
given by AdS$_2\times \RR^{2}$, whose metric and gauge field can be written as\footnote{Our gravity action is 
\be
S={1\ov 2 \kappa^2} \int  d^{4} x \ \sqrt{-g}\le[\sR +{6\ov R^2}+{R^2\ov g_F^2}F_{MN}F^{MN}\ri]\  \label{maxact}
\ee
 and we will set $g_F =1$.}
\be \label{ads2M}
ds^2 =  \mu_*^2 R^2  \le(- e^{-2 y} dt^2 + d x_1^2 + dx_2^2 \ri) +  R_2^2 dy^2  , \qquad A_t = {\mu_* \ov \sqrt{2}} e^{-y}
\ee
where $R$ and $R_2 = {R \ov \sqrt{6}}$ are 
is the curvature radius of AdS$_4$ and AdS$_2$ respectively, 
and for convenience we have introduced $\mu_* \equiv {\mu \ov \sqrt{3}}$, which will be used often below. 
The black hole horizon is located at $y=+\infty$ and the AdS$_2\times \RR^{2}$ region smoothly matches to the rest of the black hole geometry near $y \sim 0$.\footnote{Note that the constant factor $\mu_*^2 R^2$ before $dt^2$ comes from shifting the origin of $y$ so that the matching region to the rest of the black hole is near $y \sim 0$.}
 The AdS$_2$ factor, which involves the time $t$ and radial direction $y$ of the black hole, has an $SL(2,\RR)$ isometry, \ie\ the symmetry group of conformal quantum mechanics, including a scaling symmetry $t \to \lam t , y \to y + \log \lam$ acting in the time direction.
The black hole also has a nonzero entropy density of order $O(N^2)$ in the zero temperature limit. 

From the duality, the AdS$_2 \times \RR^2$ part of the charged black hole geometry maps to a nontrivial  {\it phase} of  the boundary system at finite chemical potential, to which we refer as a \slql\  for reasons to be elaborated below. 
 This phase emerges at energy scales \nb{lower than} the chemical potential $\mu$, and is characterized by  a $(0+1)$-dimensional CFT~\cite{Faulkner:2009wj}  dual to the AdS$_2$ factor, with nontrivial scaling behavior only in the {\it time} direction. More explicitly, the Fourier transform $\sO_\vk (t)$ of  a local operator $\sO(t, \vec x)$ develops a nontrivial momentum-dependent scaling dimension 
 $\de_k$ ($ k = |\vk|$), and the spectral function for $\sO$ exhibits ${\om \ov T}$ scaling
 \be \label{spec}
 A (\om, k) \propto (2 \pi T)^{2 \nu_{k}} g_1\( {\omega \over T}, \nu_k \),  \qquad 
 \nu_k \equiv \de_k - \ha 
 \ee
with $g_1$ a scaling function such that when $T \ll \om$,
\be \label{spec1}
 A (\om, \vec k) \propto \om^{2 \nu_k} \ .  
\ee 
Both $g_1$ and $\nu_k$ can be obtained explicitly from AdS$_2$ 
gravity~\cite{Faulkner:2009wj,Faulkner:2011tm}. As an example, for a charged scalar operator dual to a bulk field of charge $q$ and mass $m$, $\nu_k$ is given by
 \be \label{scaldimw}
\nu_k  = {1 \ov \sqrt{6}} \sqrt{m^2 R^2 +  { k^2 \ov \mu_*^2} - {q^2 \ov 2} + {1 \ov 4}} \ .
\ee

\subsection{Semi-local quantum liquids}


The momentum dependence in equations~\eqref{spec}--\eqref{scaldimw} has some important features. Firstly, the dependence is only through the ratio $k/\mu$, which implies that
if the momentum range one is probing the system is small compared with the chemical potential $\mu$, the scaling dimension $\de_k$ and the self-energy become approximately momentum independent.
This is reminiscent of the Marginal Fermi liquid phenomenology for cuprates~\cite{varmaetal} and the 
spin susceptibility observed for $CeCu_{6-x}Au_x$ near a antiferromagnetic critical point~\cite{shroeder}.   

Secondly, $\nu_k$, which can be rewritten as 
\be \label{defnk}
\nu_k = {1 \ov \sqrt{6} \mu_*} \sqrt{k^2 + {1 \ov \xi^2}}, \qquad \xi  
=  {1 \ov \sqrt{6} \nu_{k=0} \mu_*}
\ee
has a branch point at  $k= i \xi^{-1}$. This leads to exponential decay in spatial directions at large distances with a correlation length given by $\xi$. Generically, this correlation length is of order the inverse chemical potential, but near a quantum phase transition (such as that discussed in Section \ref{sec:sc}, $\nu_{k=0}$ goes to $0$ and the correlation length diverges. Away from such a transition, the Euclidean correlation function $G_E (\tau = i t , \vec x )$ for a scalar operator in coordinate space has two distinct regimes: 

\ben 
\item  For $x \equiv |\vec x|  \ll \xi$ (but not so small that the vacuum behavior takes over), 
\be \label{timde}
G_E (\tau, x) \sim {1 \ov \tau^{2 \de_{k=0}}} \ .
\ee

\item   For $x \gg \xi$, the correlation function decays at least 
exponentially as 
\be \label{spaco}
G_E (\tau, x) \sim e^{- {x \ov \xi}} \ .
\ee 

\begin{figure}[h]
\begin{center}
\includegraphics[scale=0.6]{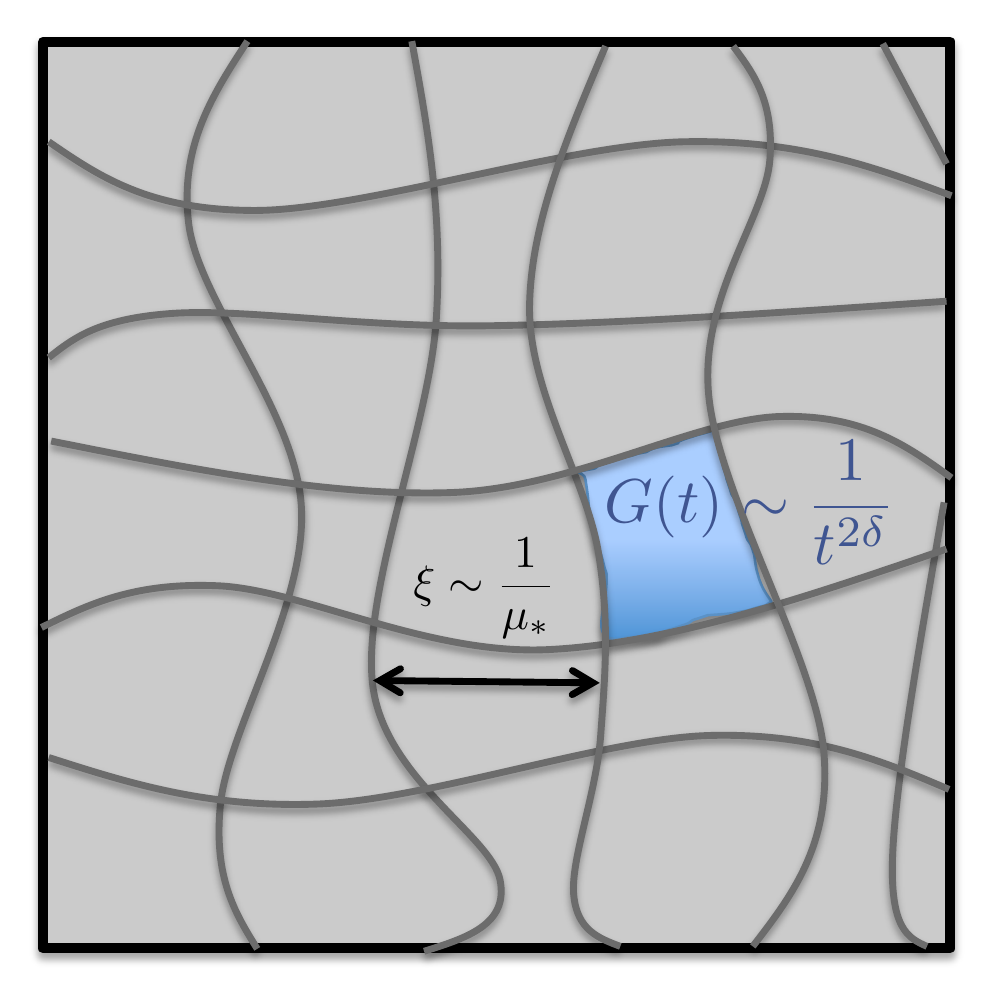}
\end{center}
\caption{Heuristic visualization of semi-local quantum liquid phase; system splits into many different weakly correlated domains, each of which is governed by a conformal quantum mechanics.}
\label{fig:semi}
\end{figure}

\een
Fig.~\ref{fig:semi} provides a heuristic visualization of the behavior: the system roughly separates into independent clusters of size of order $\xi$, with the dynamics of each cluster controlled by the $(0+1)$-d CFT. Given that the system has a nonzero entropy density, each cluster has a nonzero entropy that counts the number of degrees of freedom inside the cluster. 

While~\eqref{timde} and~\eqref{spaco} can be found by doing Fourier transforms explicitly, they can also be seen geometrically using a geodesic approximation \nb{to calculate field-theoretical correlation functions} using the Euclidean analytic continuation of~\eqref{ads2M}. 
Consider a geodesic that starts at $y \sim 0$ (which we take to be the cutoff boundary of AdS$_2$), moves into AdS$_2$ at larger values of $y$, eventually turns around, and returns to the boundary but at a spatial separation $x$ and a Euclidean temporal separation of $ \tau$. In the geodesic approximation \nb{the Euclidean correlation function} $G_E \sim e^{- m L (\tau, x)}$ where $L (\tau, x)$ is the proper distance along the geodesic. Since the metric~\eqref{ads2M} is just a direct product, we can simply find the distance moved in each factor and add them using Pythagoras. The distance moved in the $\mathbb{R}^2$ factor is $\mu_* R x$. A standard calculation shows that the distance moved in the AdS$_2$ factor is $2 R_2 \log\le(\frac{\tau}{\ep}\ri)$ where $\ep$ is an IR cutoff of the AdS$_2$,\footnote{The easiest way to understand this result is to note that its exponential must reproduce the conformal result, $\tau^{-2 m R_2}$.} which is our case is of order ${1 \ov \mu}$.
  From the IR/UV connection, an IR cutoff $\ep$ of AdS$_2$ translates into a short-distance cutoff in the \nb{infrared AdS$_2$ theory}, and thus we will restrict to $\tau > \ep$. Combining these results we find
\bea
G_E (\tau, x) &\sim & \exp\le(-m R_2 \sqrt{4 \log^2 \le(\frac{\tau}{\ep}\ri) + 6 \mu_*^2 x^2}\ri)  \cr
& \sim &  \exp \le(-\sqrt{4 \de^2  \log^2 \le(\frac{\tau}{\ep}\ri)  + {x^2 \ov \xi^2}} \ri)
 \label{geodcorr}
\eea
with $\de = mR_2$ and $\xi^{-1} = m R \mu_* $. The geodesic approximation applies to $m R \gg 1$ and in this regime, $\xi$ in~\eqref{defnk} and~\eqref{timde} and~\eqref{spaco} are indeed recovered from~\eqref{geodcorr}. 

That in~\eqref{ads2M} different points on the $\mathbb{R}^2$ can be thought of as being in different disconnected domains with size $\xi \sim {1 \ov \mu}$ can also be seen geometrically 
as follows. Consider two spacetime points on a hypersurface of given $\zeta$. To see whether observers at those locations can communicate with each other we look at {\it time-like} geodesics 
in~\eqref{ads2M} which connect the two points. Simple calculations  give that there is a maximal separation in $\mathbb{R}^2$
directions for two points to communicate with each other, given by
\be \label{maxd}
\Delta x_{\rm max} = \pi R_2  {1 \ov \mu_* R}  = \frac{\pi}{\mu_* \sqrt{6}} \ .
\ee
The first factor $\pi R_2$ in the first equality above is the time for a time geodesic to approach the boundary and come back and the second factor ${1 \ov \mu_* R}$ is the effective velocity in $\mathbb{R}^2$ (see~\eqref{ads2M}). Equation~\eqref{maxd} is consistent with $\xi$ in~\eqref{defnk} up to a prefactor.

The phase described by AdS$_2 \times \RR^2$ is also reminiscent of various theoretical models based on a large spatial dimension mean field approximation~\cite{DMFT}, such as the gapless quantum liquids of~\cite{SY,zhuetal,bgg} and the ``local quantum critical point'' of~\cite{Si.01}. 
Some suggestions for a connection have been made in~\cite{Sachdev:2010um,yamamoto,Subir2,Subir3}. 
We should emphasize that the $(0+1)$-dimensional CFT here describes, however, not the behavior of a single site, but the collective behavior of a large number of sites (if one considers our systems as a continuum limit of a lattice) over size of order $\xi$. This aspect is reflected in that, while there is nontrivial scaling only  in the time direction, the scaling dimension~\eqref{scaldimw} and correlation functions~\eqref{spec} depend nontrivially on $k$. As discussed earlier around~\eqref{defnk} it is precisely this dependence that gives the spatial correlation length of the system. 
It is also important to emphasize that, despite the scaling behavior in~\eqref{spec} and~\eqref{spec1}, we are describing a phase, not a critical point.  We thus call it a {\it \slql}, or \Slql\ for short.

Since the Einstein-Maxwell 
system is a universal sector common to many theories with a gravity dual,  the \Slql\ phase appears to be universal among a large class of field theories, independent of their microscopic details. 

\begin{figure}[h]
\begin{center}
\includegraphics[scale=0.5]{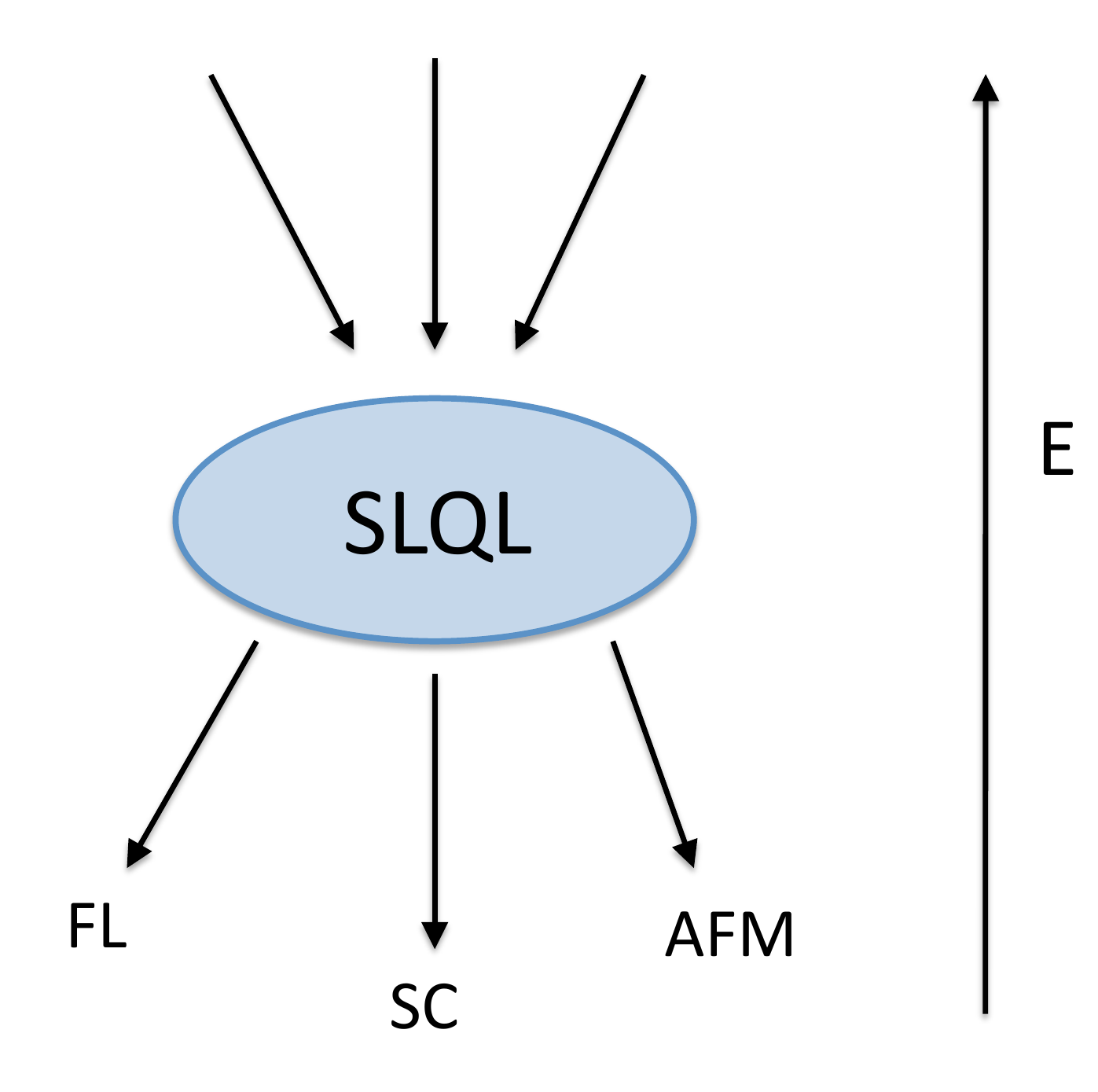}
\end{center}
\caption{The semi-local quantum liquid phase is a useful description at intermediate scales; many different UV theories (i.e. those with gravity duals described by a Maxwell-Einstein sector) flow to it, and at low energies it settles into one of many different ground states.}
\label{fig:phase}
\end{figure}

\subsection{\Slql\ as a universal intermediate-energy phase}

The nonzero entropy density of the system at $T=0$
implies that the ground states are highly degenerate. 
In a system without supersymmetry or other apparent symmetries (as it is here) to protect such degeneracies, we expect that this nonzero entropy density is likely a consequence of the $N \to \infty$ regime we are working with, i.e. this nonzero entropy density likely reflects the existence of a large number of closely spaced states which are separated from the  genuine ground state by spacings which go to zero in the $N \to \infty$ limit. One expects that the system should pick a unique ground state, which may or may not be visible in the large $N$ limit. Given that there are many nearly degenerate low energy states, the physical nature of the precise ground state should be sensitive to the parameters and the specific dynamics of an individual system. 
 Indeed, on the gravity side depending on the spectrum of charged or neutral matter fields  (their charges and masses etc.), a charged black hole suffers various bosonic and fermionic instabilities~\cite{Gubser:2008px,Hartnoll:2008vx,Hartnoll:2008kx,Iqbal:2010eh,Hartnoll:2009ns,Nakamura:2009tf}. 
The resulting state can be interpreted in the boundary theory as a superconducting/superfluid (SC) state, some analogue of an antiferromagnetic (AFM) state, or as we shall see in next section a Fermi liquid (FL) state.  The Einstein-Maxwell system can also be generalized to an Einstein-Maxwell-dilaton system in which charged black hole is no longer a solution. Instead, one finds a solution with zero entropy density and Lifshitz-type scaling in the 
interior (at $T=0$)~\cite{Gubser:2009qt,Goldstein:2009cv,Goldstein:2010aw,Charmousis:2010zz}. In all of these situations, there exist parameter ranges in which a charged black hole provides a good description for some region of the bulk geometry. 

It then appears that among a large class of systems with different microscopics {\it and} different ground states, the charged black hole appears as an intermediate energy state.\footnote{\nb{The fact that an extremal black hole should be interpreted as an intermediate-energy state has been expressed before: some earlier discussion includes~\cite{Faulkner:2009wj,Sachdev:2010um,yamamoto,mac,Faulkner:2011tm}}. In particular, see \cite{Jensen:2011su} for an argument indicating that any sort of scaling symmetry in only the time direction cannot persist to arbitrarily low energies.} Thus the \Slql\ phase  may be considered a {\it universal intermediate energy phase} which connects microscopic interactions with  macroscopic, low energy physics, as indicated in Fig.~\ref{fig:phase}. We should emphasize that for a given system there may not always exist an energy range in which \Slql\ manifests as an intermediate state. \Slql\ behavior is manifest when there exists a hierarchy between the chemical potential $\mu$ and the energy scale at which a more stable lower energy phase (say Fermi liquid or superconductor) emerges. On the gravity side there is th!
 en an intermediate region of bulk spacetime which resembles that of AdS$_2 \times \RR^2$ (or its finite temperature generalization). 
 In a situation where such a hierarchy does not exist -- for example in a holographic superconductor with transition temperature $T_c$ of order $\mu$ -- there is no temperature range over which the bulk geometry resembles that of AdS$_2 \times \RR^2$. Nevertheless, 
 as we will elaborate in more detail in Sec.~\ref{sec:sc}, the \Slql\ still provides 
 a useful description for the onset of superconductivity. 
 
As articulated by Anderson some time ago~\cite{anderson}, the existence of a ``universal intermediate phase'' appears to be a generic phenomenon in nature; the familiar examples include liquid phases of ordinary matter, through which materials settle into different crystal structures at low temperatures. 
We thus have good reasons to believe that the appearance of a \Slql\ phase in holographic systems at a finite chemical potential is not tied to the large $N$ regime we are working with, although the large $N$ limit does  magnify the universality by pushing the low energy boundary of the \Slql\ phase to zero temperature.

In the next two sections we discuss the evolution of the \Slql\ to various lower energy phases including Fermi liquids,  superconductors, and  AFM-like states. 
We argue that the emergence of these lower energy phases can be characterized as a consequence of  bound states formation in the \Slql. Conversely, \Slql\ can be considered as 
a universal deconfined (fractionalized) phase of these lower energy phases.

\section{Fermi liquids from a \Slql\ } \label{sec:fermi}

In this section we show that in a certain parameter regime a Fermi liquid phase of heavy fermions emerges from the \Slql\ phase through formation of fermionic bound states.

Consider a bulk fermionic field $\psi$ of charge $q$ dual to some fermionic composite operator $\sO$ in the boundary system. 
The conformal dimension $\De$ of $\sO$ in the vacuum is related to the mass $m$ of $\psi$ by $\De ={3 \ov 2}  + m R$
where $R$ is the AdS curvature radius. In the infinite $N$ limit, at a finite chemical potential $\mu$  the retarded function $G_R (\om, \vk)$ of $\sO$ at low energies $\om \ll \mu$
was found earlier by solving the bulk Dirac equation in the
charged black hole geometry~\cite{Lee:2008xf,Liu:2009dm,Cubrovic:2009ye,Faulkner:2009wj}, i.e. in the \Slql\ phase. Depending on the value of $q$ and $m$, the fermionic excitations exhibit a variety of non-Fermi liquid behavior~\cite{Liu:2009dm,Faulkner:2009wj,Faulkner:2010da,Faulkner:2011tm}. In particular,  
for $q^2 > 2 m^2 R^2$, the fermionic operator in momentum space $\sO_\vk (t)$ develops a {\it complex} scaling dimension $\de_k = \ha + i \lam_k$ 
in the \Slql\ for $k < k_o$ where 
\be 
k_o = 
\mu_* \sqrt{{q^2 \ov 2} - m^2 R^2} , \qquad \lam_k =  
{1 \ov \sqrt{6} \mu_*} \sqrt{ k_o^2 - k^2} 
\ .
\ee
The corresponding retarded function exhibits oscillatory behavior in $\log \om$ and is given by
\be
G_R(\om, k) =  
h(k)  \frac{e^{i \th_b (k)}  +e^{- i \th_b (k)}  c(\lam_k) \le(\frac{\om}{\mu_*}\ri)^{-2i\lam_k}}{e^{i \th_a (k)} + e^{- i \th_a (k)}c(\lam_k) \le(\frac{\om}{\mu_*}\ri)^{-2i\lam_k}} \label{oscGr}
\ee
where $h(k), \th_a (k), \th_b (k)$ are real functions of $k$ which can be found numerically while $c(\lam_k)$  is a complex function whose form is explicitly known. The $\om \to 0$ limit of~\eqref{oscGr} is singular and at $\om =0$, the spectral function has a non-vanishing weight for $k < k_o$  
\be 
\Im G_R(\om = 0, k) = 
h(k) \sin (\th_b (k)- \th_a (k))  \  .
\label{specF}
\ee
Equation~\eqref{specF} indicates the presence of gapless excitations, and indeed for any $k < k_o$~\eqref{oscGr} has an infinite number of poles in the lower complex $\om$-plane which accumulate at the origin. But there are no long-lived quasiparticles as all these poles have comparable real and imaginary parts.  
When $k > k_o$, the spectral function is identically zero\footnote{Note in~\eqref{specF} as $k \to k_o$, $\th_a (k)$ and $\th_b (k)$ become equal.}  at $\om =0$, but exhibits scaling behavior for nonzero $\om$ 
\be
\Im G_R(\om, k)  
\propto \om^{2 \nu_k}  , \qquad  \nu_{k} = {1 \ov \sqrt{6} \mu_*} \sqrt{k^2 - k_o^2} \ .
\label{specF1}
\ee
If the fermion operator further satisfies  $q^2 > 3m^2 R^2$, then there exist additional  Fermi surfaces with Fermi momentum $k_F > k_o$ and the retarded function near such a $k_F$ is given by ($c(\nu_{k_F})$ is complex)
\be \label{NFLF}
G_R (\om, k) = \frac{h_1}{k-k_F- \frac{1}{v_F}\om - \Sigma(\om)}, \quad \Sig (\om) \propto c(\nu_{k_F}) \om^{2 \nu_{k_F}}, \quad \nu_{k_F} = {1 \ov \sqrt{6} \mu_*} \sqrt{k_F^2 - k_o^2} \ .
\ee
The exponent $\nu_{k_F}$ is real and controls whether or not the Fermi surface has long-lived quasi-particles~\cite{Faulkner:2009wj}. When $m^2R^2 > {q^2 \ov 2}$, then $k_o^2 <0$, for which there is no oscillatory behavior or isolated Fermi surfaces, and  one simply finds scaling behavior~\eqref{specF1} for any $k$.

\nb{The above discussion is for the standard quantization of a fermion in the bulk. For $mR \in (0, \ha)$, there also exists an alternative quantization for which $\De = {3 \ov 2} - mR$. In the alternative quantization there exists an isolated Fermi surface with behavior~\eqref{NFLF} even for $q^2 < 2 m^2 R^2$. }

\begin{figure}[h]
\begin{center}
\includegraphics[scale=0.7]{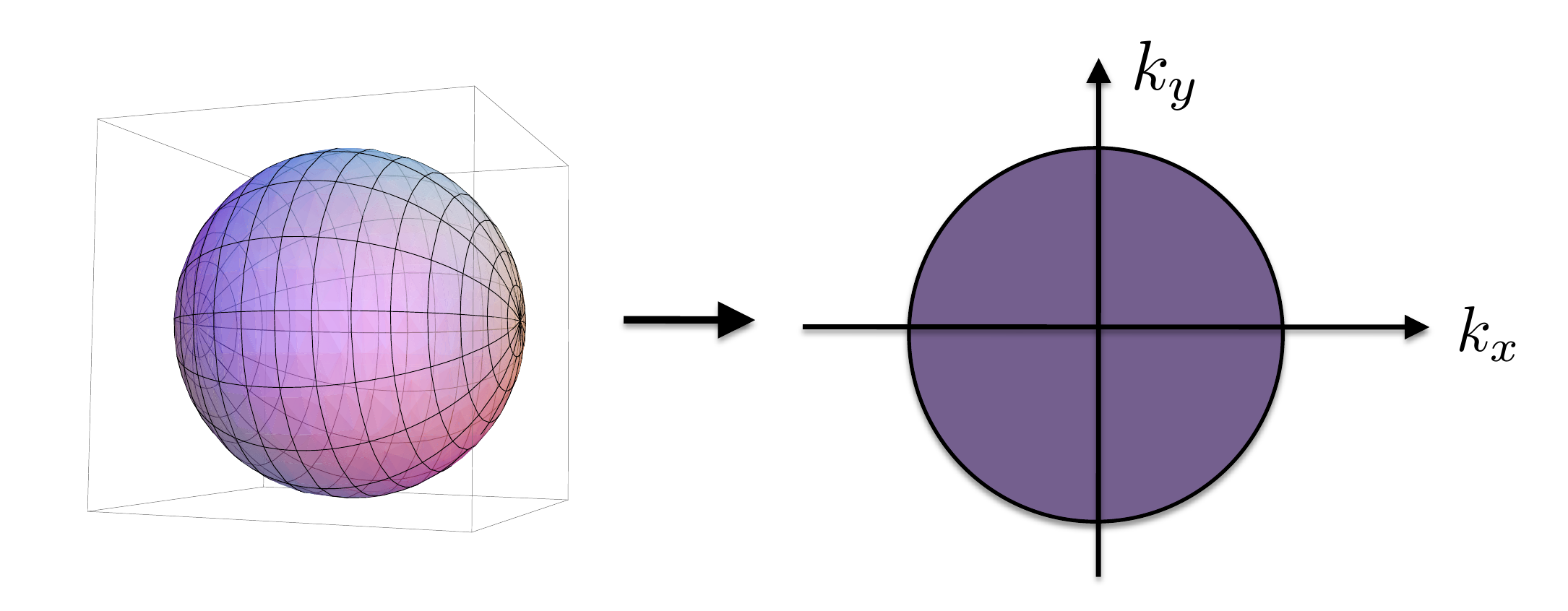}
\end{center}
\caption{In the AdS$_2 \times \RR^2$ region of the extremal black hole geometry,  at each point in the bulk there is a local three-dimensional Fermi surface with Fermi momentum $k_o$, which upon projection to the boundary theory would result in a Fermi disc, in which there are gapless excitations at each point in the {\it interior} of a disc in the two-dimensional momentum space. }
\label{fig:bal}
\end{figure}

\begin{figure}[h]
\begin{center}
\includegraphics[scale=0.55]{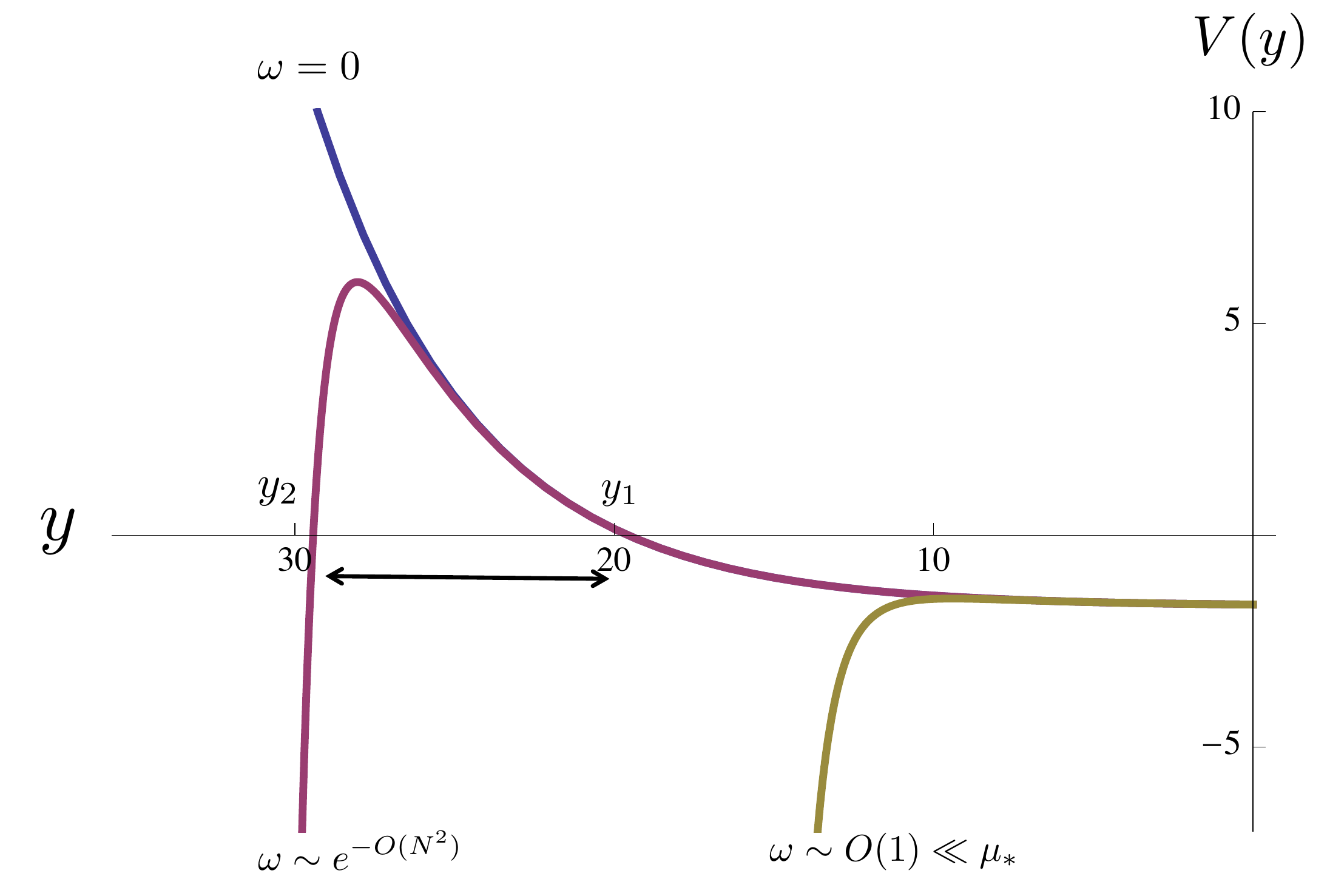}
\end{center}
\caption{Plots of the WKB potential~\eqref{potm} for various values of $\om$ in the Lifshitz region. 
For $\om =0$, the potential is bounded and the Bohr-Sommerfeld quantization gives normalizable 
modes in the bulk which correspond to different Fermi surfaces in the boundary theory. For an exponentially small $\om$, although the potential is unbounded from below in the large $y$ region, the excitations have a small imaginary part due to the potential barrier. For $\om_c \ll \om  \ll \mu$ (with $\om_c$ given by~\eqref{2co}), the potential barrier disappears and the potential becomes the same as that for the AdS$_2 \times \RR^2$. }
\label{fig:wkbP}
\end{figure}

The above results are for the $N \to \infty$ limit. We now show that for $q^2 > 2 m^2 R^2$ and a large but finite $N$, at very low energies the system is in fact a Fermi liquid with a large number ($O(N^2)$) of densely spaced Fermi surfaces.  For this purpose, let us start with a heuristic explanation of the nonzero spectral weight~\eqref{specF} from the black hole geometry. 
When $q^2 > 2 m^2 R^2$, quanta for $\psi$ can be pair produced by the electric field of the black hole~\cite{Pioline:2005pf}, which
results in a gas of $\psi$ quanta hovering outside the horizon. The ground state of this fermionic gas is described by a bulk Fermi surface which can be shown to have a Fermi momentum given precisely by $k_o$ by using the Thomas-Fermi approximation discussion in Sec. 7.4 of~\cite{Hartnoll:2009ns}.  The projection of this bulk Fermi surface to the boundary then leads to gapless excitations for all $k < k_o$. See Fig.~\ref{fig:bal}. The small excitations at the bulk Fermi surface have a large decay rate since they can fall into the black hole without experiencing any potential barrier (see also Fig.~\ref{fig:wkbP}). 

\begin{figure}[h]
\begin{center}
\includegraphics[scale=0.5]{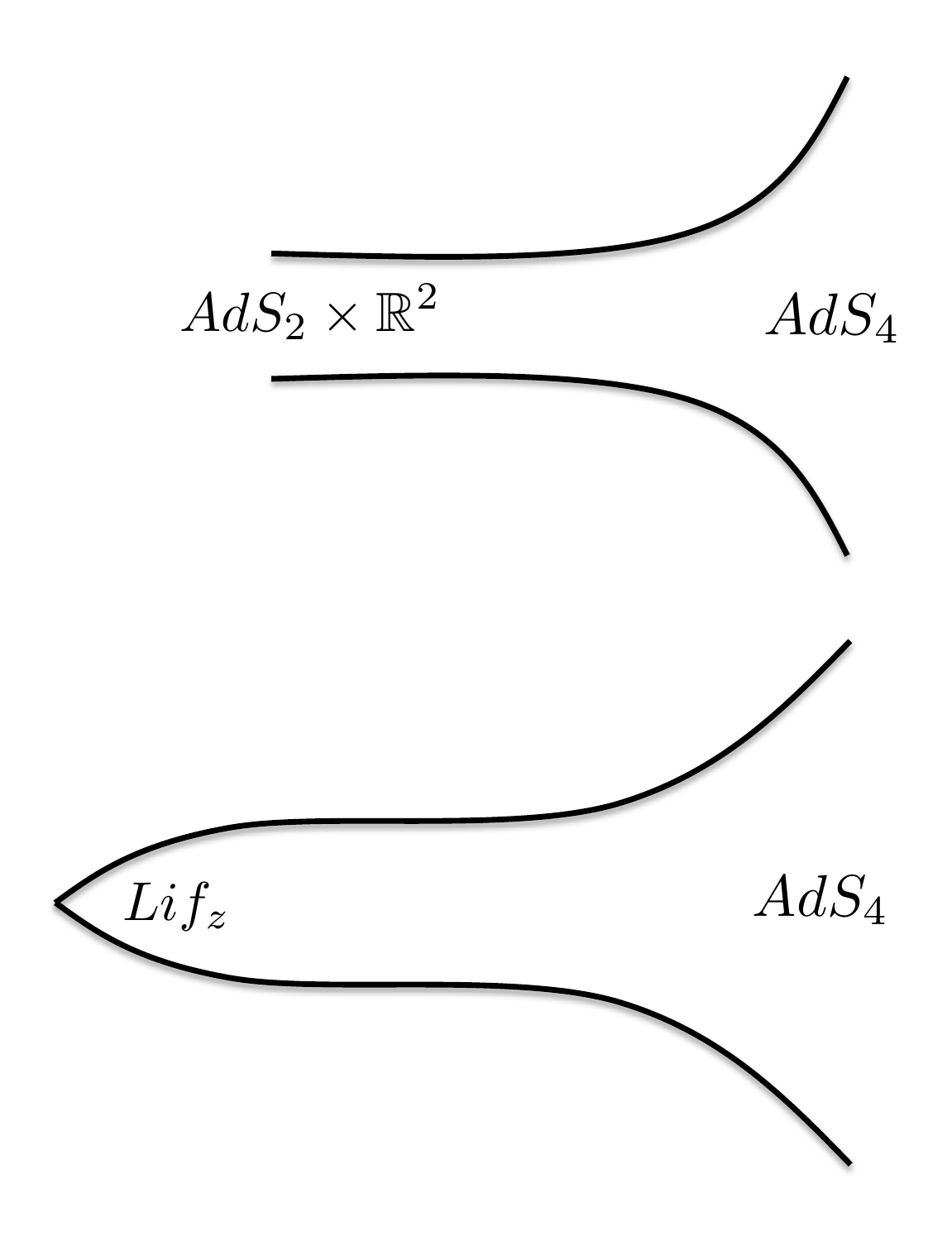}
\end{center}
\caption{Two different geometries: on the top, the AdS$_2 \times \mathbb{R}^2$ describing the SLQL phase; on the bottom, its resolution into a Lifshitz geometry with a finite $z$ given by \eqref{zexp}. The horizon direction represents the $y$ direction, while the vertical direction represents the transverse $\RR^2$ (i.e. $x_1, x_2$ directions). In the plot for the Lifshitz geometry it should be understood that the tip lies at an infinite proper distance away. When $z$ is large as in~\eqref{zexp}, there is a large range of $y$ for which the Lifshitz geometry resembles that of  AdS$_2 \times \mathbb{R}^2$. Also note that $e^{-y}$ translates into the boundary theory energy scale.
}
\label{fig:comparison}
\end{figure}

The charge density carried by the black hole
is determined by the classical geometry and background fields, giving rise to a boundary theory charge density
\be \label{toach}
\rho_0 = {R^2 \ov G_N} {\mu^2 \ov 4 \sqrt{3} \pi} \sim O(N^2) \ .
\ee 
In contrast the local bulk density for $\psi$ quanta produced from quantum pair production gives a contribution of order $ O(N^0)$, which naively implies that the fermionic backreaction can be ignored at leading order. 
However, 
since in the near horizon region~\eqref{ads2M} the proper radial distance is infinite and the local proper volume of the transverse $\RR^2$ is constant, the total boundary density coming from integrating over the radial direction is in fact infinite~\cite{Hartnoll:2009ns}. 
The backreaction of the fermionic gas is thus important, after taking into account of which, the near-horizon geometry is modified from~\eqref{ads2M} to~\cite{Hartnoll:2009ns} 
\be \label{lifM}
ds^2 = \mu_*^2 R^2  \le(- e^{-2y} dt^2 + e^{-{2 y \ov z}} (dx_1^2 + dx_2^2)  \ri) + R_2^2 d y^2 , \qquad A_t = {\mu_* \ov \sqrt{2}} e^{-y}
\ee
with 
\be  \label{zexp}
{1 \ov z} =  {2 q \ov \sqrt{6} \pi} {G_N \ov R^2} \le({k_o \ov \mu} \ri)^3 \sim {1 \ov N^2} \ .
\ee
In the backreacted geometry~\eqref{lifM}, shown schematically in Figure \ref{fig:comparison}, the local proper volume of the transverse $\RR^2$ goes to zero as $y \to \infty$, resulting in a  finite boundary fermionic density of order $O(N^2)$, given by
\be \label{boude}
\rho_F = {z q k_o^3 \ov 6 \sqrt{2} \pi^2 \mu} \ . 
\ee
Plugging the explicit value~\eqref{zexp} of $z$ into~\eqref{boude} we find that~\eqref{boude} becomes identical to~\eqref{toach}, i.e.  all the charge density of the system is now carried by the fermioinic gas; the black hole has disappeared! The system now also has zero entropy density. Note that in obtaining~\eqref{zexp} and~\eqref{boude}, one used the Thomas-Fermi approximation which is valid when $q$ and $mR$ are taken to be parametrically large. 

Now consider the Dirac equation for the bulk fermions in the backreacted geometry~\eqref{lifM}, which 
for  $q$ and $mR$ parametrically large can be conveniently solved using the WKB approximation and describes a particle of {\it zero energy} moving in a potential 
\be \label{potm}
V (y) = {1 \ov 2 \mu^2} \le(k^2 e^{2y \ov z} - k_o^2 - \om^2 e^{2y} - \sqrt{2} \mu_* q \om e^y \ri) \  . 
\ee
Since $z \gg 1$,  the metric~\eqref{lifM} and~\eqref{potm} can be well approximated by that of AdS$_2 \times \RR^2$ until  $y  \gtrsim z \sim O(N^2)$.\footnote{\nb{The analysis of this section overlaps with the recent WKB work of \cite{Hartnoll:2011dm}; see also the discussion at the end of Section \ref{sec:con}.}}
 
This implies that the boundary physics will deviate from that described by the \Slql\ (i.e. a charged black hole) results described earlier only at exponentially small energies. More precisely, for $k < k_o$, one finds that 
for  
\be \label{2co}
\om  \ll \om_c (k) \equiv {k_o \ov z } \exp \le(- z \log {k_o \ov k}\ri)
\ee
the boundary retarded Green function is modified from~\eqref{oscGr} to 
\be \label{furet}
G_R (\om,k) 
=
h (k) {\cos (\th - \th_b) -{i \Ga \ov 4} \sin (\th-\th_b)  \ov 
\cos (\th - \th_a) -{i \Ga \ov 4} \sin (\th-\th_a)} + \cdots\ . 
\ee
where $h(k)$, $\th_a, \th_b$ are the same functions as in~\eqref{oscGr}--\eqref{specF}. 
$\th$ and $\Ga$ are complicated {\it real} function of $k$ and $\om$, given by
\be \label{verdf} 
\th = z f(k) + C(k) \om + \cdots , \qquad \Ga \sim  \exp \le(- {\sqrt{2} z k \ov \mu} \le({k \ov  \om} \ri)^{1 \ov z} \ri)
\ee
where 
\be 
f(k) = {k_o \ov \sqrt{2} \mu}  \le[\log \le({k_o \ov k}  + \sqrt{{k_o^2  \ov k^2} -1  } \ri) - { \sqrt{1 - {k^2 \ov k_o^2} }}\ri]
, \quad C(k) = {\sqrt{\pi z} q  \ov 2 \sqrt{6} k_o} e^{z \log {k_o \ov k}}  \ .
\ee
In the above expressions we have only kept the leading order terms in $1/z$ and $1/q$.

\begin{figure}[h]
\begin{center}
\includegraphics[scale=0.8]{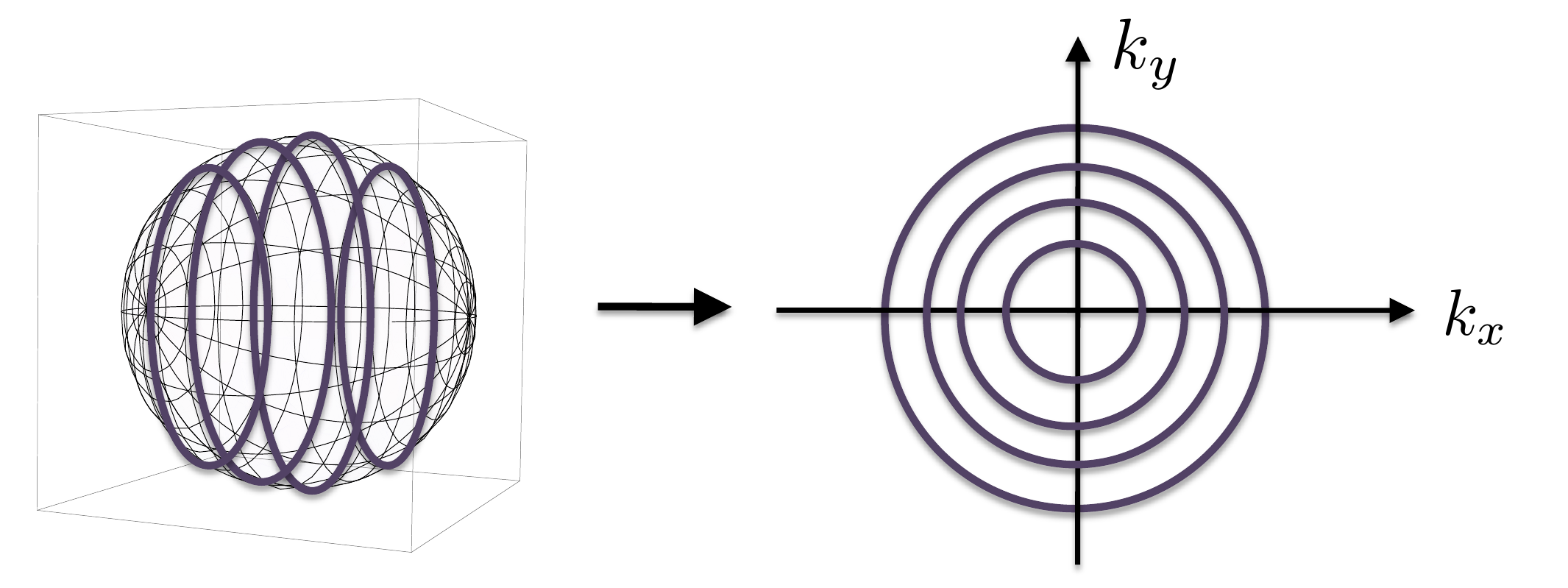}
\end{center}
\caption{Taking into account the quantization condition~\eqref{roro}, we find instead a set of discrete states in the bulk in which radial motion is quantized. This results in a family of concentric Fermi surfaces in the boundary theory, which resolves the Fermi disk of Fig.~\ref{fig:bal}.}
\label{fig:concentric}
\end{figure}

The poles of~\eqref{furet} give a family of densely packed Fermi surfaces with Fermi momenta given by
\be \label{roro}
z f(k_F^{(n)}) - \th_a (k_F^{(n)})=\le(n+\ha \ri) \pi
\ee
with the largest Fermi momentum of order 
\be  \label{kfma}
k_F^{(max)} \approx k_o - O\le(z^{-{2 \ov 3}} \ri) \approx  k_o - O\le(N^{-{4 \ov 3}} \ri) \ .
\ee
See Figure ~\ref{fig:concentric}. The spacings between nearby Fermi momenta are of order $N^{-{4 \ov 3}}$ near $k_F^{(max)}$, but become $O(N^{-2})$ when $k_F^{(n)}$ is $O(1)$ from $k_o$.\footnote{The equation~\eqref{roro} itself appears to imply an infinite number of Fermi surfaces with spacing of order ${k_F \ov z}$ as $k_F \to 0$. Likely one cannot trust the analysis when $k_F$ becomes some inverse power of $N$.}
Near any such $k_F$, one finds that 
\be \label{retg}
G_R (\om, k) = 
h (k_F)  {\sin (\th_a - \th_b)   \ov 
C (k_F) \om + z f'(k_F) (k-k_F) + {i \Ga \ov 4} } 
\ee
with
\be \label{deri}
f'(k) = - {1 \ov \sqrt{2} \mu k } \sqrt{k_o^2 - k^2} \ 
\ee
which can be written in a canonical way as 
\be 
G_R (\om, k) = - {Z  \ov 
 \om - v_F (k-k_F) + \Sig}
\ee
with
\be \label{ferP}
Z = \le(h (k) C^{-1} (k) \sin (\th_b - \th_a) \ri)_{k=k_F}, \quad v_F = - C^{-1} (k_F)  z f'(k_F), \quad
\Sig = {i \ov 4} C^{-1} (k_F)  \Ga (k_F) \ .
\ee
For $\om=0$, equation~\eqref{retg} recovers~\eqref{specF} in the limit $z \to \infty$. 
The Fermi velocity for quasiparticle excitations $v_F  \propto \sqrt{z} e^{-z \log {k_o \ov k}}$ is exponentially small, indicating an exponentially large effective mass. These are heavy fermions! Also note that the decay rate which is proportional to $\Sig$ is exponentially small both in $z$ and $1/\om^z$, as anticipated earlier in~\cite{Faulkner:2010tq}. The quasiparticle weight $Z$ can also be written as 
\be \label{pepm}
Z (k)= C^{-1} (k) \Im G_R^{N = \infty} (\om=0, k)
\ee
where we have used~\eqref{specF}. The first factor in~\eqref{pepm}, proportional to $ e^{-z \log {k_o \ov k}}$, is exponentially small and increases with $k$, while the second factor (independent of $z$),  known numerically~\cite{Liu:2009dm},  is a decreasing function of $k$ and approaches zero at $k_o$. Since $z$ is very large here, we expect $Z$ should increases with $k$ all the way to $k^{max}_F$.\footnote{The maximum of~\eqref{pepm} is reached when $k = k_o - O(1/z)$ which is larger than $k_F^{max}$ in the large $z$ limit.}  

We now see that the oscillatory region $k < k_o$ splits into a large number of closely spaced Fermi surfaces with associated quasiparticles having a very large effective mass and a very small decay rate. In this regime the Luttinger theorem should hold, which we now check explicitly. The total boundary volume of the Fermi surfaces is given by
\bea 
V_{FS} &= & \pi  \sum_n  (k_{F}^{(n)})^2  = \pi \sum_n \int  dk \,  \delta (k-k_n) k^{2}   \cr
&=&  {z } \sum_n {\pi \ov z} \int d k \, |f'(k)| \delta \le(f(k) -{n \pi \ov z} \ri) k^{2}
\eea
The last line can now be approximated by a continuous integral for a large $z \sim O(N^2)$ leading to 
\be 
V_{FS} =  {z}    \int_0^{k_o} d k \, |f'(k)|  k^{2} 
= {z k_o^3\ov 3 \sqrt{2} \mu} 
\ee
where we have used~\eqref{deri}.
The Luttinger theorem then says that the charge density of the system should be
\be 
\rho_F = {2 q \ov (2 \pi)^{2}} V_{FS} = {q z k_o^3\ov 6 \sqrt{2} \pi^2 \mu}
\ee
which precisely agrees with~\eqref{boude}. Note the factor $2$ in the first equality comes from the number of helicities of the boundary fermion. 

\begin{figure}[h]
\begin{center}
\includegraphics[scale=0.55]{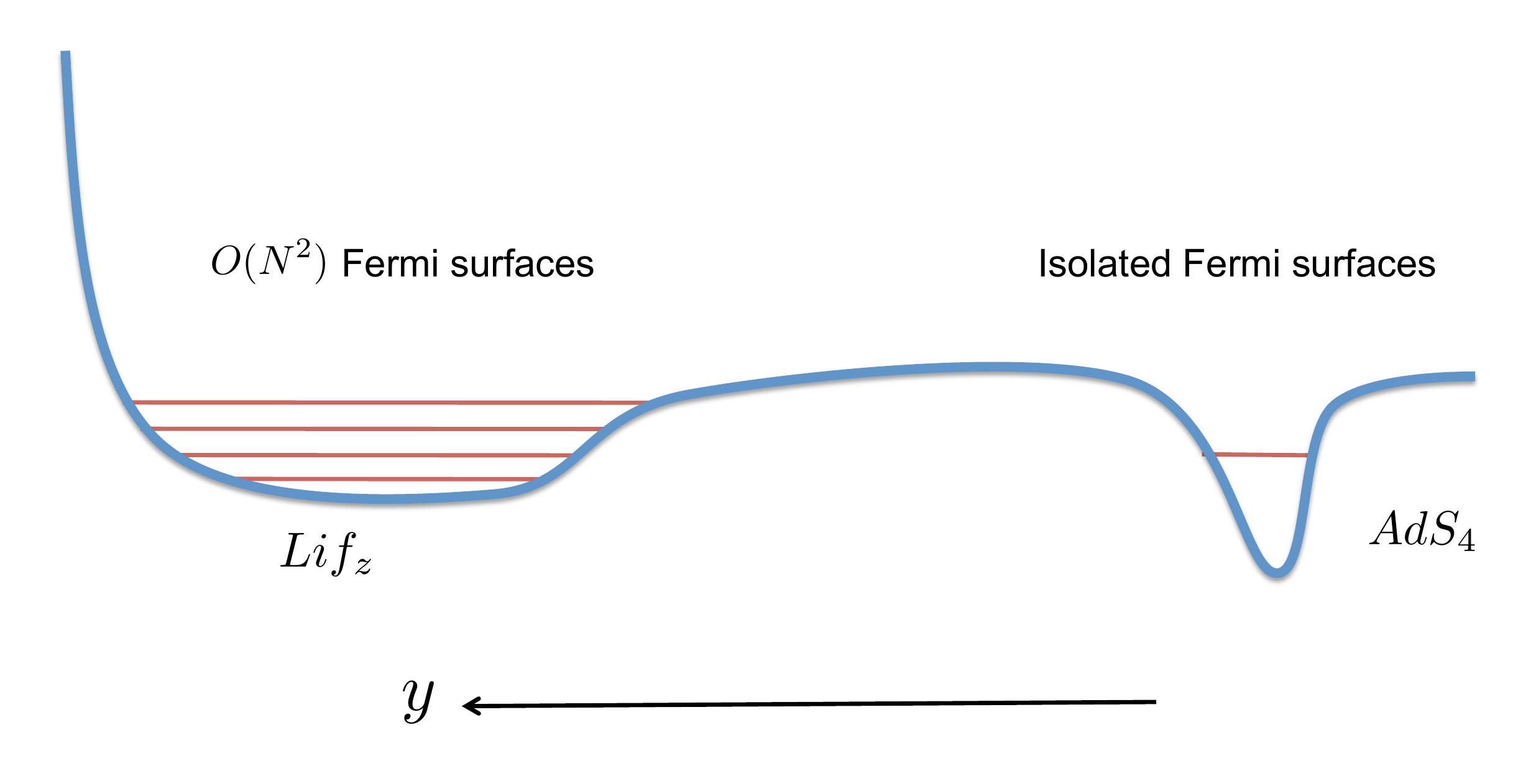}
\end{center}
\caption{A cartoon of the bound-state structure for the full spacetime, i.e. including the asymptotic AdS$_4$ region. The Fermi surfaces~\eqref{NFLF} appear as bound states in a potential well in the UV region. }
\label{fig:wkbPF}
\end{figure}

At a heuristic level, the above results can be readily understood from the qualitative features of the WKB potential~\eqref{potm}, as shown in Fig.~\ref{fig:wkbP}. In particular, the locations of the family of 
boundary Fermi surfaces for $k<k_o$ simply follow from the Bohr-Sommerfeld quantization condition for a particle moving in the potential, and the quasiparticle decay rate is given by the tunneling rate of a particle through the potential barrier. When $\om$ is large enough, i.e. outside the range~\eqref{2co}, the potential barrier disappears and one recovers the non-Fermi liquid behavior~\eqref{oscGr}--\eqref{NFLF} of a \Slql. Note that the WKB potential~\eqref{potm} only includes the near the horizon region of the backreacted charged black hole geometry, and  does not include the  
Fermi surfaces in~\eqref{NFLF}.  In Fig.~\ref{fig:wkbPF} we also show a cartoon of the WKB potential for $\om=0$ for the full spacetime, i.e. including the asymptotic AdS$_4$ region. 
The \nb{isolated} Fermi surfaces~\eqref{NFLF} appear as bound states in a potential well outside the 
near-horizon region.  

Generalizing the above analysis for general $k$, we find for $k > k_o$, at exponentially small $\om$, i.e. 
\be \label{3co}
\om \ll {k_o \ov z} \exp \le(- z \log {k \ov \sqrt{k^2 - k_o^2}} \ri)
\ee
the scaling behavior in the spectral function~\eqref{specF1} should be 
 replaced by $\Im G_R (\om, k) \propto \Ga(\om, k)$ with $\Ga$ given by~\eqref{verdf}.
Similarly for the Fermi surfaces~\eqref{NFLF} (which exist for $q^2 > 3m^2 R^2$) with $k_F > k_o$, the decay rate of the small excitations near a Fermi surface becomes exponentially small in $1/\om^z$ when $\om$ lies in the range~\eqref{3co}.  
But the Fermi velocity of quasiparticle excitations for these Fermi surfaces remains the same as in~\eqref{NFLF}, i.e. $v_F \sim O(1)$; these are light fermions. Note that since there is only an $O(N^0)$
of such Fermi surfaces each of size $O(N^0)$, at order $O(N^2)$ they do not affect the Luttinger count described earlier. 

We also note that our results only depend on $z$ being parametrically large, not necessarily of order $O(N^2)$. Thus it should also apply to a generic Lifshitz geometry for with a large $z$. 


Our procedure of solving the Dirac equation in the backreacted geometry~\eqref{lifM} is self-consistent and extracts the leading non-perturbative behavior in $1/N^2$. 
There are also perturbative loop corrections in the bulk, which give perturbative corrections in $1/N^2$ to the self-energy.  We expect the qualitative features of our results (e.g the family of densely spaced Fermi surfaces etc) to be robust against these corrections, as they have to do with the global structure of the backreacted geometry~\eqref{lifM}.
 Near the Fermi surface perturbative corrections to the self-energy should give rise to a term ${c \ov N^2} \om^2$ with $c$ some complex $O(1)$ coefficient, which will dominate over the ${i \Ga \ov 4}$ term in~\eqref{retg} for $\om$ in the range~\eqref{2co}. 
Thus we expect that the quasiparticle decay rate should be proportional to $\om^2$ as in a Landau Fermi liquid. Nevertheless, the much smaller non-perturbative correction proportional to $\Ga$ does signal some nontrivial underlying physics beyond that of a Landau Fermi liquid. Similar arguments apply to~\eqref{NFLF}; for $\om$ in the range~\eqref{3co}, the imaginary part of $\Sig$ should be proportional to $\om^2$. 

To summarize, imagine a system with only a single fermionic operator $\sO$ satisfying the condition $q^2 > 2 m^2 R^2$, and no other instabilities. Then at very low energies, the system is described by a Fermi liquid with $O(N^2)$ densely spaced Fermi surfaces, each of size $O(N^0)$. 
The quasiparticle excitations have a very large effective mass (proportional to $e^{N^2}$). 
When $q^2 > 3 m^2 R^2$ there could be some additional isolated Fermi surfaces with an $O(1)$ effective mass. At small but not exponentially small frequencies, there is a wide energy range over which the system is controlled by the \Slql, with scaling behavior in the time direction and various non-Fermi liquid behavior~\eqref{oscGr}--\eqref{NFLF}.  In the \Slql\ phase the clue that the system will eventually settles into a Fermi liquid state is the existence of a region $k < k_o$, where the scaling dimensions for fermions become complex.

We now give a physical interpretation in the boundary theory for the emergence of the Fermi liquid phase from \Slql. It is tempting to interpret different Fermi surfaces in the family of densely spaced Fermi surfaces as corresponding to {\it different bound states} generated by the fermionic operator $\sO$. 
This appears natural: as discussed earlier each of them corresponds to a different radial mode in the bulk. In particular, from equation~\eqref{2co},  Fermi surfaces of different Fermi momenta disappear at  different energy scales given roughly by $\om_c (k_F)$, with Fermi surfaces of larger Fermi momenta more stable.  Note that from~\eqref{roro}  a mode with larger $k_F$ has a smaller $n$ (i.e. with a smaller number of radial oscillations), which is consistent with general expectations. 
When turning on the temperature, each Fermi surface is destroyed  individually at a temperature scale $T_c (k_F) \sim \om_c (k_F)$ with the  largest Fermi surface being destroyed at a temperature scale 
\be  \label{tmax}
T_c^{\rm max} \sim O(e^{- z^{1 \ov 3}}) \sim O(e^{- N^{2 \ov 3}}) \ .
\ee
 In other words, as we increase the temperature, there exist a hierarchy of scales at which each bound state of $\sO$ is 
``ionized'' individually. 

Recall that in our set-up, $\sO$ is a composite operator of fundamental fields. The number of degrees of freedom for the fundamental fields is $O(N^2)$. Thus the Fermi liquid state can be considered a ``confined'' state, in which the low energy degrees of freedom are Fermi surfaces from a discrete set of composite fermionic bound states.  In contrast, the \Slql\ is a ``deconfined'' state in which the composite bound states deconfine and  fractionalize into  more fundamental degrees of freedom. 
In the bulk, the emergence of the fractionalized \Slql\ phase is reflected in the 
emergence of a charged black hole description. In the Fermi liquid state, the system is characterized by ``heavy'' fermions. But such coherent quasiparticles disappear in the \Slql. Instead one finds some kind of quantum soup which is characterized by scaling behavior~\eqref{spec}--\eqref{spec1} for any 
 bosonic or fermionic operators. 
 In our discussion the presence of  a large number of Fermi surfaces has to do with a spectrum of densely spaced bound states. The fractionalized picture should be independent of this feature.
Also note that our interpretation of~\Slql\ as a fractionalized phase resonates with an earlier discussion of~\cite{Sachdev:2010um}, although the details are different.

It is natural that in this deconfined state, there is a finite entropy density, proportional to the number of fundamental degrees of freedom of the system. From the discussion of previous paragraphs, the fractionalization and release of entropy from the Fermi liquid appear to happen individually for each Fermi surface. We can estimate the entropy associated with {\it a} Fermi surface at the temperature scale of its demise as 
\be 
\de s = \int^{T_c (k_F)}_0 dT {C_v \ov T}   \sim {1 \ov z^{3 \ov 2}}
\ee
where $C_v \propto {k_F \ov v_F} T $ is the specific heat associated with the Fermi surface and 
$T_c (k) \sim \om_c (k)$ is determined by~\eqref{2co}. 
Note that due to large effective masses of quasiparticles, the Fermi liquid phase has an exponentially large specific heat $C_v \propto e^{z \log {k_o \ov k}}$ and the entropy of the system increases over an exponentially small temperature range $T_c (k)$  to an order only power suppressed in $z$.  

While a Lifshitz geometry certainly does not describe a confined phase for the full $(2+1)$-dimensional boundary theory, in addition to the entropy count, there is another important aspect that a Lifshitz geometry can be considered as a confined state of \Slql\ represented by AdS$_2 \times \RR^2$. A defining feature of \Slql\   is that at small $\om$ the spectral weight for a generic operator scales with $\om$ as a power for any momentum $k$ (see equations~\eqref{spec}--\eqref{spec1}), which indicates the presence of a large number of low energy excitations for all momenta (at larger momenta the weight will be suppressed by a higher power).  In the Lifshitz geometry, generalizing our earlier WKB discussion to generic operators, one finds that (as in~\eqref{furet} and~\eqref{verdf})
\be \label{onesp}
A (\om, k) \propto  \exp \le(- {\sqrt{2} z k \ov \mu} \le({k \ov  \om} \ri)^{1 \ov z} \ri)
\ee
i.e. the spectral weight becomes exponentially suppressed in $1/\om$ at small frequencies. While~\eqref{onesp} indicates there are still some low energy excitations remaining, most low energy excitations in AdS$_2 \times \RR^2$ have disappeared at sufficiently low energies.

The isolated Fermi surfaces described by~\eqref{NFLF} exist both in the Fermi liquid phase and in the
\Slql, i.e. such fermionic excitations remain confined in the deconfined \Slql\ phase. The excitations are Landau quasiparticles in the Fermi liquid phase, but become non-Fermi liquids (with or without well-defined quasiparticles) from hybridization with the deconfined fermionic degrees of freedom in the \Slql~\cite{Faulkner:2009wj,Faulkner:2010tq,Sachdev:2010um}. Note that there is a parameter range for $2 m^2 R^2 < q < 3 m^2 R^2$ in which the Fermi liquid phase exists, but there are no isolated Fermi surfaces. 

Finally we note that the large hierarchy of scales between the chemical potential $\mu$ and
the scale~\eqref{tmax} at which the \Slql\ completely sets in should be attributable to the fact that the fermionic 
bound states in our system have very small binding energies. It would be interesting to find examples with larger binding energies. The ``electron star'' geometry discussed in~\cite{Hartnoll:2010gu, Hartnoll:2010xj}  (see also~\cite{Arsiwalla:2010bt,vCubrovic:2010bf}) could be such candidates if a proper context can be found to give the $N$-scaling there a sensible boundary theory interpretation. 


\section{Scalar condensates from a \Slql\ } \label{sec:sc}

In this section we consider scalar instabilities of a \Slql \;phase. 
We show that a confinement mechanism similar to that underlying the transition to a Fermi liquid phase applies to scalars as well. This gives an underlying physical mechanism for a large class of holographic superconductors discussed in the literature~\cite{Hartnoll:2009sz,herzogr,Horowitz:2010gk}. While 
at a finite density a fermionic bound state forms a Fermi surface, a scalar bound state forms a Bose-Einstein condensate. 

Consider a scalar operator $\sO$ of charge $q$  dual to a bulk scalar field $\phi$ of mass $m$. The vacuum conformal dimension $\De$ of $\sO$ is related to the mass $m$ of $\phi$ by 
$\De = {3\ov 2} + \sqrt{m^2 R^2 + {9 \ov 4}}$. 
When $\sO$ is charged, its condensate can be interpreted as a  superconductor (or more precisely a charged superfluid). For a neutral $\sO$ (i.e. $q=0$),  as discussed in~\cite{Iqbal:2010eh} the condensed phase can be used as a model for antiferromagnetism when it is embedded as part of a triplet transforming under a global $SU(2)$ symmetry corresponding to spin. For a single real $\sO$ with a $Z_2$ symmetry, the condensed phase can be used as a model for an Ising-nematic phase. Below we will simply refer to the condensate of a neutral $\sO$ as an AFM-type state.

In the \Slql\ the scalar operator $\sO_{\vk} (t)$ in momentum space develops a new scaling dimension given by equation~\eqref{scaldimw}. The system is unstable to forming condensate of $\sO$ 
when the scaling dimension is complex at $k=0$~\cite{Faulkner:2009wj}, i.e.  
when 
\be 
q^2 > 2 m^2 R^2 + {1 \ov 2} 
\ee
which on the gravity side corresponds to violating the Breitenlohner-Freedman bound~\cite{bf} of AdS$_2$~\cite{Gubser:2005ih,Hartnoll:2008kx,Gubser:2008pf,Denef:2009tp}.
At a finite temperature, the instability sets in at some critical temperature $T_c$,  where the system transitions to a superconducting or an AFM-type state. If $T_c \ll \mu$, then there is a temperature range $T_c \leq T \ll \mu$ in which the system is described by a \Slql\ with ${\om \ov T}$ scaling as in~\eqref{spec}. In the gravity description, there is a region of spacetime which resembles a black hole in AdS$_2 \times \RR^2$. When $T_c$ becomes comparable to $\mu$, such an intermediate regime is not manifest. Nevertheless, as we argue below, the \Slql\ remains a useful description for understanding the origin of the instability. In particular, similar to the fermionic case, we find that the developing of a complex scaling dimension for a scalar operator $\sO$ in \Slql\ signals formation of bound states of $\sO$.

Consider a ``thought'' experiment in which we could dial the vacuum dimension $\De$ (or the infrared mass $m$ of the corresponding bulk field) using some external knob.\footnote{Whether one can really do so in a specific model is not important for our purposes. One could view this as a mathematical device to elucidate the nature of the instability. We do point out that various holographic constructions do provide us with such a knob \cite{Iqbal:2010eh,Jensen:2010ga,Jensen:2010vx}, often related to an applied magnetic field.} We could then smoothly tune $T_c$ 
to zero temperature at a ``quantum critical point'' given by $m_c^2 R^2 = {q^2 \ov 2} - {1 \ov 4}$. 
For $m^2 > m_c^2$ the instability disappears.  Near $m_c^2$ (on the unstable side), there is a wide range of energy scales where the system is described by \Slql\ and 
one has the advantage that the onset of the instability can be studied analytically in terms of the AdS$_2$ geometry. Interestingly one finds that the analysis in AdS$_2$ becomes almost identical~\cite{Kaplan:2009kr,Jensen:2010ga,Iqbal:2010eh} to that of the formation of three-body bound states 
in the Efimov effect~\cite{efimov}. In particular, one finds a tower of geometrically spaced Efimov states, which Bose condense with an expectation value~\cite{Iqbal:2010eh} 
\be
\vev{\sO}_n \sim \mu_*^\De \exp \le(- {\sqrt{6} n \pi \ov R \sqrt{m_c^2 - m^2}} \ri), \qquad 
n=1,2,\cdots \ .
\ee
The new vacuum is given by the condensate of the lowest state ($n=1$). As in the Efimov effect, 
here the continuous scaling symmetry (in the time direction) of \Slql\ is broken to a discrete one due to complex scaling dimension. When we move deeper into the unstable region by decreasing $m^2$, 
the scaling symmetry of \Slql\ is less manifest and it becomes harder to identify the excited Efimov states. Nevertheless, the above physical picture should still apply.

The physical picture here is similar to the BEC regime in a strongly interacting ultracold Fermi system 
where fermions form bound molecules and then Bose condense. In such a situation one expects to see also an intermediate regime in which the system forms bound molecules, but not yet Bose condense. This intermediate stage has not been explicitly identified in the gravity side.  It would be interesting to investigate it further. There is also another channel for scalar instabilities in a charged black hole geometry first found in~\cite{Faulkner:2009wj} from existence of certain inhomogeneous scalar black hole hair and further generalized and studied in~\cite{Faulkner:2010gj} using double trace deformations. The physical mechanism underlying this instability does not involve bound state formation and is analogous to the BCS regime of a strongly interacting ultracold Fermi system. 

We conclude this section by pointing out an important difference between a holographic superconductor and an AFM-type state. Similarly to the fermionic case, the geometry for a holographic superconductor at zero temperature is given by a Lifshitz geometry (which includes AdS$_4$ as a special example)~\cite{Gubser:2009cg,Horowitz:2009ij,g1} in the infrared. Such a solution is believed to be stable, and as discussed around~\eqref{onesp} in Sec.~\ref{sec:fermi}, in a Lifshitz geometry almost all the low energy excitations  present at generic momenta in an AdS$_2 \times \RR^2$ are gapped out. In contrast, the infrared region of the bulk geometry for the condensate of a neutral scalar 
is still given by an AdS$_2 \times \RR^2$, but with a smaller curvature radius and entropy density than those of the uncondensed geometry~\cite{Iqbal:2010eh,Faulkner:2010gj}. This implies that such a neutral condensate 
is not yet the stable ground state, and at even lower energy 
 some other order has to take over. For example if there are fermionic operators in the system satisfying $q^2 > 2 m^2 R^2$, then the system will eventually settle into a Fermi liquid phase coexisting with the AFM-type order.

\section{Discussion} \label{sec:con}

In this paper we argued that gauge/gravity duality applied to strongly interacting finite-density systems predicts a universal intermediate-energy \Slql\ phase. In the \Slql\ the system is characterized by a nontrivial scaling behavior in the time direction (and associated $\om/T$ scaling) and has a finite entropy density. At lower energies, the \Slql\ may order in many different ways to other more stable phases, depending on the specific details of an individual system (there are a whole string landscape of possibilities!). We considered three classes of such lower energy phases: a Fermi liquid phase of heavy fermions, a superconducting state, and an AFM-type state. 
The common threads among them are that certain operators develop a complex scaling dimension 
in the \Slql, which then leads to the formation of bound states of the operator at lower energies. For a  fermionic operator, the bound states form Fermi surfaces, while for a scalar operator, the bound states Bose condense. Conversely, the \Slql\ can be considered a deconfined phase where composite bound states fractionalize into more fundamental degrees of freedom. 

It would be interesting to search for other instabilities of the \Slql\ (i.e. instabilities of a charged black hole), which could help identify new low energy phases. 

 An important conclusion of our results is that the
holographic (non)-Fermi liquids discussed in recent literature~\cite{Lee:2008xf,Liu:2009dm,Cubrovic:2009ye}, which correspond to the isolated Fermi surfaces described in Sec.~\ref{sec:fermi}, are consequences of intermediate-energy effects, rather than reflecting properties of the true ground state. The motivates the further search for holographic non-Fermi liquids in gravity systems which extend down to arbitrarily low energies and so are related to the structure of the ground state (for a recent attempt, see e.g.~\cite{Iizuka:2011hg}).  

Our discussion also clarifies the physical mechanism through which a class of holographic superconductors arise from \Slql\ through bound state formation. It would be interesting to investigate further whether one can identity a regime where bound states have formed, but have not yet Bose condensed. 

It is tempting ask whether the \Slql\ phase identified here could underlie some known strongly correlated condensed matter systems. When one encounters scaling behavior in an observable, 
an important immediate question is whether the behavior is due to intermediate-energy or vacuum effects.  
A prominent example is the  linear temperature dependence of the low temperature resistivity occurring in the normal state of cuprate superconductors near optimal doping and heavy electron systems near a critical point. While a quantum critical point naturally gives rise to 
scaling behavior, examples of an intermediate phase with scaling behavior have been hard to come by in higher than $(1+1)$-dimension.\footnote{In $1+1$-dimensions, an example is the spin-incoherent Luttinger liquid (see e.g.~\cite{fiete}). We thank T.~Senthil for this suggestion.}
Thus the \Slql\  could provide a useful reference point for such a discussion. 

In heavy electron systems, while quantum fluctuations from the quantum critical point corresponding to the onset of magnetism provide a natural starting point for understanding the observed non-Fermi liquid behavior, 
there appears plenty of room for certain non-Fermi liquid behavior to arise from an intermediate-energy phase. 
We note that (see e.g. ~\cite{stewart}): 

\ben

\item In various 
experiments the non-Fermi-liquid behavior often extends to temperatures and parameter regions 
 far away from the critical point. There also exist materials in
which the non-Fermi-liquid behavior exists in the magnetically ordered phase.

\item Non-fermi liquid phases which are not associated with a quantum critical point have also been 
 observed. (Such non-Fermi liquid behavior could also be due to some stable spin liquid phase.)

\item   
Inelastic neutron-scattering experiments have seen
critical behavior at a range of wave vectors not close to the ordering wave vector.  

\item While many heavy-fermion materials
display non-Fermi-liquid physics in the vicinity of the
onset of magnetism,  the detailed behavior near a quantum critical point 
appears to differ significantly among different systems. In particular, in
some materials the superconducting state could dominate over 
 the magnetic state at low temperatures.

\een
Thus it would be important to distinguish the non-Fermi liquid behavior which arises from intermediate-energy effects from that due to a quantum critical point. 
Also in many heavy electron materials, the appearance of novel superconductivity has often been associated with the onset of magnetism. It would be interesting to explore whether some of them can in fact be attributed to an intermediate-energy phase such as the \Slql. 

Similarly one could also question whether the strange metal behavior and the onset of superconductivity in cuprates could be consequences of some intermediate-energy phase like the \Slql. In particular, in~\cite{Faulkner:2009wj,Faulkner:2010da} some intriguing parallels between  holographic non-Fermi liquids and the strange metal phase of cuprates have been found.\footnote{Note that in curprates the linear resistivity behavior has been measured to very low temperatures, which could be an indication that in curprates the strange metal behavior is due to vacuum properties rather than intermediate energy effects.}

More generally, we expect that candidates for the \Slql\ to occur include  
systems which exhibit frustrated or competing interaction terms in their Hamiltonian. Such systems can have a large number of near-degenerate states near the vacuum, similarly to the holographic systems considered here. Also systems which involve strong competition between tendencies towards itinerancy and localization could exhibit the semi-local behavior found here.
 There are also some tantalizing parallels between properties of the \Slql\ phase and models from dynamical mean field theory (DMFT)~\cite{DMFT}, such as scalings in time direction with spatial directions as spectators and a finite entropy. 
 The fact that DMFT techniques have been very successful in treating many materials in some intermediate energy region suggests 
that there could be an underlying universal intermediate-energy phase like \Slql. 

%

Finally we mention that the confining mechanism through which Fermi liquids of heavy fermions arise from \Slql\ may also result in new ideas for 
understanding heavy electron systems. Theoretical models for heavy electrons in the slave boson formalism have largely been based on the Higgs mechanism.  It would be interesting to find new 
models that use confinement instead.  

\medskip

\centerline{\it Note added:}

 While this paper was being finalized,~\cite{Hartnoll:2011dm} appeared which overlaps with our 
WKB analysis in Sec.~\ref{sec:fermi} of fermions in the backreacted charged black hole geometry,   including the presence of a large number of densely spaced Fermi surfaces which satisfy the Luttinger theorem.~\cite{Hartnoll:2011dm} works in the ``electron star'' geometry~\cite{Hartnoll:2010gu, Hartnoll:2010xj} which is also characterized by an IR Lifshitz geometry, but with $z \sim O(1)$, for which there is not an intermediate AdS$_2$ region. As in our discussion of holographic superconductors in Sec.~\ref{sec:sc}, one can imagine smoothly dialing the parameters of the ``electron star'' so that $z$ becomes parametrically large. In the large $z$ regime, there is again an intermediate \Slql\ phase and the lower energy state with multiple Fermi surfaces can be interpreted as a confined state of the \Slql. 
When $z$ is not parametrically large, while the \Slql\ is no longer manifest, it still provides a useful way for understanding the emergence of multiple Fermi surfaces at low energies. 

We understand that a related investigation also appears in \cite{new}. 


\vspace{0.2in}   \centerline{\bf{Acknowledgements}} \vspace{0.2in} We thank P.~Coleman,  T.~Faulkner, S.~Hartnoll, D.~Hofman, G.~Kotliar,
 P.~Lee, J.~McGreevy, D.~Park, J.~Polchinski, S.~Sachdev,  K.~Schalm, Q.~Si, D.~T.~Son, D.~Vegh, X-G.~Wen, and J.~Zaanen for illuminating discussions, and in particular T.~Senthil for various long and instructive discussions. Work supported in part by funds provided by the U.S. Department of Energy
(D.O.E.) under cooperative research agreement DE-FG0205ER41360.

\end{document}